\documentclass[aps,pre,twocolumn,superscriptaddress,showpacs]{revtex4-1}

\usepackage{amsmath,amssymb,bm}
\usepackage{graphics}
\usepackage{epsfig}
\usepackage{graphicx}

\newcommand{\mean}[1]{\left\langle#1\right\rangle}

\begin{document}

\title{Kuramoto model with frequency-degree correlations on complex networks
}

\author{B.~C. Coutinho}
\affiliation{Departamento de F{\'\i}sica da Universidade de Aveiro, I3N, 3810-193 Aveiro, Portugal}
\affiliation{Center for Complex Networks Research, Northeastern University, Boston, Massachusetts 02115, USA}

\author{A.~V. Goltsev}
\affiliation{Departamento de F{\'\i}sica da Universidade de Aveiro, I3N, 3810-193 Aveiro, Portugal}
\affiliation{A.F. Ioffe Physico-Technical Institute, 194021 St. Petersburg, Russia}

\author{S.~N. Dorogovtsev}
\affiliation{Departamento de F{\'\i}sica da Universidade de Aveiro, I3N, 3810-193 Aveiro, Portugal}
\affiliation{A.F. Ioffe Physico-Technical Institute, 194021 St. Petersburg, Russia}

\author{J.~F.~F. Mendes}
\affiliation{Departamento de F{\'\i}sica da Universidade de Aveiro, I3N, 3810-193 Aveiro, Portugal}

\date{\today}

\begin{abstract}
We study the Kuramoto model on complex networks, in which natural frequencies of phase oscillators and the vertex degrees are correlated. Using the annealed network approximation and numerical simulations we explore a special case in which the natural frequencies of the oscillators and the vertex degrees are linearly coupled. We find that in uncorrelated scale-free networks with the degree distribution exponent $2 < \gamma < 3$, the model undergoes a first-order phase transition,
while the transition becomes of the second order at $\gamma>3$. If $\gamma=3$, the phase synchronization emerges as a result of a hybrid phase transition that combines an abrupt
emergence of synchronization, as in first-order phase transitions, and a critical singularity,
as in second-order phase transitions. The critical fluctuations manifest themselves as avalanches in synchronization process. Comparing our analytical calculations with numerical simulations for Erd\H{o}s--R\'{e}nyi and scale-free networks, we demonstrate that the annealed network approach is accurate if the the mean degree and size of the network are sufficiently large. We also study analytically and numerically the Kuramoto model on star graphs and find that if the natural frequency of the central oscillator is sufficiently large in comparison to the average frequency of its neighbors, then synchronization emerges as a result of a first-order phase transition. This
shows that oscillators sitting at hubs in a network may generate a discontinuous synchronization transition.
\end{abstract}
\maketitle


\section{Introduction}



%

Synchronization phenomena
attracted  much attention of the scientific community in the last decades but the understanding of emergence of synchronization in complex systems is still an open problem \cite{Pikovsky}.  A few examples are the flashing of fireflies, the chirp of the crickets, the pacemaker cells of the heart, and synchronous neural activity. The Kuramoto model \cite{Kuramoto75,Kuramoto84} stands out as the classical paradigm for studying spontaneous emergence of collective synchronization in complex systems (see, for example, Refs.~\cite{arenas,Acebron2005,Dorogovtsev:dgm08}). This basic model is
analytically treatable and may contribute to
general understanding of  synchronization phenomena.

The Kuramoto model describes a system
of interacting phase oscillators. An explicit solution of this model was found for an infinite complete graph with a symmetric single peaked distribution of natural frequencies and an uniform coupling constant $J$ \cite{Kuramoto75,Kuramoto84,Strogatz1991}. In this case, when the coupling between oscillators becomes greater than a critical value $J_c$, the spontaneous synchronization emerges as a result of a second-order phase transition
with the standard mean-field critical exponent $\beta=1/2$ for the order parameter.
Further investigations demonstrated, however, that the kind of the phase transition depends on the form of the distribution
of the natural frequencies of the oscillators.
The Kuramoto model with a convex distribution function undergoes a discontinuous transition in contrast to the second-order transition with $\beta=1/2$ when the distribution function is concave \cite{Acebron2005}. In the particular case of a flat distribution of natural frequencies, the synchronization emerges discontinuously as a result of the hybrid phase transition
with a jump of
the order parameter as in a first-order phase transition, but
also with strong critical fluctuations as in a continuous
phase transition
\cite{Pazo2005,bu2007,bu2008}.

Many real-world complex systems have a structure of random complex networks \cite{ab2001,dg2002,newman2003}, and this kind of structure can strongly influence their dynamics \cite{Dorogovtsev:dgm08}. Within the Kuramoto model, the structure of the underlying network also plays an important role and affects the synchronization of oscillators. The Kuramoto model with a symmetric single peaked  distribution function of natural frequencies on uncorrelated random scale-free complex networks with a degree distribution $p(q)\propto q^{-\gamma}$ was studied in works \cite{Ichinomiya2004,Ichinomiya2005,Lee2005} by use of a mean-field approach.
It was shown that if
the second moment of the degree distribution is finite in the infinite size limit (i.e., at $\gamma >3$), then
the critical coupling $J_c$ is finite and the phase transition is of the second order. In contrast to this kind of networks, in networks with a diverging second moment (i.e., at $2 < \gamma \leq 3$), the critical coupling $J_c$ tends to zero in the infinite size limit. This means that an arbitrary finite coupling leads to synchronization of phase oscillators. A similar critical properties were found for the Ising and Potts models on scale-free networks \cite{dgm2002,dgm2004,Dorogovtsev:dgm08}.

Recently, an interesting variation of the Kuramoto model was proposed by G\'{o}mez-Garde\~nes et al. \cite{Gomez_2011}. The authors introduced a model
in which natural frequencies $\omega_j$ and degrees $q_j$ of vertices are rigorously (namely, linearly) related, $\omega_j=a q_j + b$.
By use of numerical simulations of the model with $N=1000$ oscillators, they found that a second-order phase transition occurs in Erd\H{o}s--R\'{e}nyi networks and in the model \cite{gm2006}.
In the configurational model of scale-free networks with $\gamma < 3.3$ and in the Barab\'{a}si-Albert model they observed a first-order phase transition at
a finite critical coupling $J_c$
in contrast to the zero critical coupling found in Ref.~\cite{Ichinomiya2004,Ichinomiya2005,Lee2005} for $\gamma < 3$ in the infinite size limit. G\'{o}mez-Garde\~nes et al. suggested \cite{Gomez_2011} that this discontinuous transition may be driven by hubs that entrain and synchronize neighboring oscillators. Recently,
Leyva et al. \cite{Leyva2012} showed that the first-order phase transition is not specific for the Kuramoto model, but also can be found in other systems of non-linear oscillators, for example, in  scale-free networks of interacting piecewise R\"{o}ssler units.
The transition was confirmed experimentally in electronic circuits with a star graph configuration \cite{Leyva2012}. The fact that this first-order phase transitions can be observed experimentally opens ground for technological application and makes the
understanding of this behavior even  more urgent.

In the present paper, in order to understand the role of
frequency-degree correlations for synchronization of phase oscillators,
we carry out a detailed
analysis of the model proposed by G\'{o}mez-Garde\~nes et al. \cite{Gomez_2011} in the case of networks with scale-free topology.  Using
the annealed network approach \cite{Dorogovtsev:dgm08,Bianconi} and performing numerical simulations of the model, we show that the model actually undergoes a first-order phase synchronization transition
if the underlying networks have scale-free network topology with the degree distribution exponent $2<\gamma < 3$. For scales-free networks with $\gamma>3$ the system demonstrates a second-order phase transition with the mean-field critical exponent $\beta=1/2$ for the order parameter. Surprisingly, we find a hybrid phase transition with $\beta=2/3$ at $\gamma=3$. In the latter case, synchronization emerges discontinuously with increasing coupling between oscillators but hysteresis is absent and there are critical fluctuations as at second-order phase transitions. Interestingly, these critical phenomena are related to the avalanches of synchronization between oscillators.
Furthermore, in order to understand a role of hubs for synchronization, we also study the Kuramoto model on star graphs and find a criterion for the first-order synchronization transition.

\section{General equations}
\label{Generalequations}


 The dynamics of phase oscillators in the Kuramoto model is described by
the following equations:
\begin{equation}
\dot{\theta}_j=\omega_j+ \sum_{l=1}^{N} J_{j l} a_{j l}\sin(\theta_l-\theta_j)
,
\label{eq1}
\end{equation}
where $N$ is the total number of oscillators, $\theta_j$ and $\omega_{j}$ are, respectively, the phase and the natural frequency of
oscillator $j$, where
$j=1,...,N$. $J_{j l}>0$ is the coupling between oscillators $j$ and $l$. $\dot{\theta_j}$ is defined as $\dot{\theta_j}\equiv d\theta_j/dt$. $a_{j l}$ is the entry of the adjacency matrix of the network. $a_{j l}$ is equal to $1$ if vertices $j$ and $l$ are connected, and $a_{j l}=0$ if they are not. For simplicity, we assume that the coupling constant is uniform, i.e., $J_{j l}=J$. 

Let us use the annealed network
approximation to solve this model on an uncorrelated random complex network \cite{Dorogovtsev:dgm08,Bianconi}. Within this approach,
the entries $a_{j l}$ in Eq.~(\ref{eq1}) are replaced by the probabilities $a^{(an)}_{j l}$,
\begin{equation}
a^{(an)}_{j l}= q_j q_l/(N \langle q \rangle),
\label{annealed_network}
\end{equation}
that vertices $j$ and $l$ with degrees $q_j$ and $q_l$, respectively, are connected.
$\langle q \rangle$ is the mean degree,
$\langle q \rangle\equiv \sum_{j} q_j/N$.
Here the annealed network approximation plays the role of a mean-field approach.
Substitution of Eq.~(\ref{annealed_network}) to Eq.~(\ref{eq1}) means that
the actual interactions of phase oscillators with their nearest neighbors are replaced by weighed interactions with all of the oscillators.
As a result, Eq. (\ref{eq1}) takes a form,
\begin{equation}
\dot{\theta}_j-\Omega=\omega_j-\Omega-J r q_j \sin(\theta_j-\psi),
\label{eq3}
\end{equation}
where
\begin{equation}
r e^{i \psi} \equiv \frac{1}{N \langle q \rangle}\sum_{j=1}^{N} q_j e^{i \theta_j}.
\label{kuramoto3}
\end{equation}
The parameter $r$ is the order parameter of synchronization, $\psi$ represents a global phase of the system, and $\Omega$ is the group angular velocity, $\Omega\equiv \dot{\Psi}$.
We assume that in the limit $t\rightarrow+\infty$ and $N\rightarrow+\infty$, the system approaches a steady state with a constant group angular velocity $\Omega$, i.e., $\dot{\Omega}=0$.

Analyzing Eq. (\ref{eq3}), one finds that there are two groups of phase oscillators. If $|\omega_j-\Omega|<J r q_j$, then oscillator $j$ is locked. In this case, Eq. (\ref{eq3}) has a stable solution with $\dot{\theta}_j=\Omega$ and takes the form
\begin{equation}
\omega_j-\Omega=J r q_j \sin\left(\theta_j-\psi\right).
\label{stablestate}
\end{equation}
The locked oscillators are synchronized and are rotating together with the same group angular velocity $\Omega$.
If $\left.|\omega_j-\Omega|>J r q_j\right.$, oscillator $j$ is drifting and never reaches a steady
state, in contrast to the locked oscillators.

Let us study the Kuramoto model with a linear
relation between natural frequencies $\omega_j$ and degrees $q_j$ (frequency-degree correlations \cite{Gomez_2011}), i.e.,
\begin{equation}
\omega_j=a q_j+b.
\label{linear}
\end{equation}
Using a rotating frame, $\omega_j\rightarrow \omega_j- b$, and rescaling the coupling constant, $J\rightarrow J/|a|$, one obtains that the model with an arbitrary parameters $a$ and $b$ is equivalent to the model with $b=0$ and $a=1$. It is the case that we will study below. Taking into account locked and drifting oscillators, we write Eq. (\ref{kuramoto3}) as follows,
\begin{eqnarray}
\nonumber
r&=&\frac{1}{N \langle q \rangle} \sum_{j=1}^{N} q_j e^{i(\theta_j-\psi)}\Theta\left(1-\left|\frac{\omega_i-\Omega}{J r q_j}\right|\right)
\\[5pt]
&+&\frac{1}{N \langle q \rangle}  \sum_{j=1}^{N} q_j e^{i(\theta_j-\psi)} \Theta\left(\left|\frac{\omega_i-\Omega}{J r q_j}\right|-1\right),
\label{eq4}
\end{eqnarray}
where $\Theta(x)$ is the Heaviside step function. The first term is the contribution of locked oscillators to the order parameter and the second term is the contribution of drifting oscillators. 
Replacing the summation over degrees by integration and using an explicit solution of Eq.~(\ref{stablestate}), we obtain that the contribution of the locked oscillators to the order parameter in the thermodynamic limit is
\begin{eqnarray}
\nonumber
&&\frac{1}{N \langle q \rangle} \sum_{j=1}^{N} q_j e^{i(\theta_j-\psi)}\Theta\left(1-\left|\frac{\omega_j-\Omega}{J r q_j}\right|\right)\simeq
\\[5pt]
\nonumber
&&\int_{1}^{+\infty} dq\,p(q) q\sqrt{1-\left(\frac{q-\Omega}{J r q}\right)^2}\Theta\left(1-\left|\frac{q-\Omega}{J r q}\right|\right)
\\[5pt]
&+&i\int_{1}^{+\infty} dq\,p(q) \frac{q-\Omega}{J r}\Theta\left(1-\left|\frac{q-\Omega}{J r}\right|\right).
\label{lockedcontribution}
\end{eqnarray}
In the thermodynamic limit $N\to\infty$, the contribution of the drifting oscillators to the order parameter is
\begin{eqnarray}
\nonumber
&&\frac{1}{N \langle q \rangle} \sum_{j=1}^{N} q_j e^{i(\theta_j-\psi)}\Theta\left(\left|\frac{\omega_j-\Omega}{J r q_j}\right|-1\right)\simeq
\\[5pt]
\nonumber
&&i\int_{1}^{+\infty} dq\,p(q) \frac{q-\Omega}{J r}\left[1-\sqrt{1-\left(\frac{J r q}{q-\Omega}\right)^2}\,\,\right]\times
\\[5pt]
&&\Theta\left(1-\left|\frac{J r q}{q-\Omega}\right|\right).
\label{driftingcontribution}
\end{eqnarray}
(see Appendix~\ref{anexA} and \cite{Strogatz1991,bu2008}). In order to simplify our calculations, it is convenient to  introduce a variable
\begin{equation}
\alpha\equiv r J.
\label{r-alpha}
\end{equation}
Then, substituting  Eqs.~(\ref{lockedcontribution}) and (\ref{driftingcontribution}) into Eq.~(\ref{eq4}) and
considering the real and imaginary parts of the order parameter $r$, we obtain a set of two equations,
\begin{eqnarray}
&&\nonumber \langle q \rangle-\Omega= \int_{1}^{+\infty} dq \, p(q) (q-\Omega) \times \\ &&\sqrt{1-\left(\frac{\alpha q }{q-\Omega}\right)^2}\Theta \left(1-\left|\frac{q \alpha}{q-\Omega}\right|\right),
\label{eq6a}
\\[5pt]
&&R(\alpha)=\frac{\alpha}{J},
\label{eq6c}
\end{eqnarray}
for two unknown parameters $\Omega$ and $\alpha$. It is convenient to consider $\Omega$ as a function of $\alpha$, $\Omega = \Omega(\alpha)$. The function $R(\alpha)$ in Eq.~(\ref{eq6a}) is defined as follows,
\begin{eqnarray}
&& \nonumber R(\alpha)\equiv\frac{1}{\langle q \rangle} \int_{1}^{+\infty} dq \, p(q) q \sqrt{1-\left(\frac{q-\Omega(\alpha)}{\alpha q }\right)^2} \times
\\[5pt]
&& \Theta \left(\left|\frac{q \alpha}{q-\Omega(\alpha)}\right|-1\right).\label{eq6b}
\end{eqnarray}
Solving Eqs.~(\ref{eq6a}) and (\ref{eq6c}), we find $\alpha$ and the group angular velocity $\Omega$. Then, from Eq.~(\ref{r-alpha}), we find the order parameter $r$.


\section{Kuramoto model on the Erd\H{o}s--R\'{e}nyi networks}



\begin{figure}[!htb]
\centering
\includegraphics[width=0.48 \textwidth]{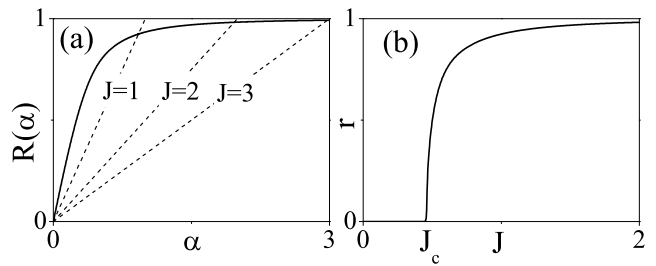}
\caption{Kuramoto model with frequency-degree correlations on Erd\H{o}s--R\'{e}nyi networks. (a) The function $R(\alpha)$ versus $\alpha$ from Eq.~(\ref{eq6b}) for the Erd\H{o}s--R\'{e}nyi network with $\langle q \rangle=10$. The dashed lines display the line $\alpha/J$ at different couplings $J$. The intersection of $R(\alpha)$ and a dashed line gives a solution of Eq.~(\ref{eq6c}) and, in turn, from Eq.~(\ref{r-alpha}), determines the order parameter $r$. (b) The order parameter $r$ versus $J$ for the network.
\label{graph1}}
\end{figure}


Let us consider the Kuramoto model with frequency-degree correlations,
 $\omega_j=q_j$, on the Erd\H{o}s--R\'{e}nyi
graph with a given mean degree $\langle q \rangle$. In this case, the degree distribution is Poissonian, $p(q)= \langle q \rangle^q e^{-\langle q \rangle}/q!$.
The function $R(\alpha)$ given by Eq.~(\ref{eq6b})
is represented in Fig.~\ref{graph1}(a). Solving numerically Eqs.~(\ref{eq6a}) and (\ref{eq6c}), we find that a non-trivial solution appears if $J$ is greater than a critical coupling $J_c$. The order parameter $r$ as a function of $J$ is
shown in Fig.~\ref{graph1}(b). Expanding the function $R(\alpha)$ at $\alpha\ll1$, we find the critical behavior of $r$ near $J_c$,
\begin{equation}
r\propto (J-J_c)^{\beta},
\label{eq7}
\end{equation}
where the critical exponent $\beta$ equals $1/2$ and the critical coupling $J_c$ is
\begin{equation}
J_c=\frac{2 \langle q \rangle}{\pi p\left(\Omega\right)\Omega^2}.
\label{eq8}
\end{equation}
At the critical point $J=J_c$, the group angular velocity is $\Omega=\Omega(\alpha=0)$ and can be found from the equation
\begin{equation}
\int_1^{+\infty}dq p(q) \frac{q^2}{q-\Omega}=0.
\label{eq9}
\end{equation}
At $\langle q \rangle\gg 1$, equations~(\ref{eq8}) and (\ref{eq9}) give
\begin{equation}
J_c = 2\sqrt{2}/ \sqrt{ \pi \langle q \rangle}.
\label{eq10}
\end{equation}
Note that this asymptotic result has a square-root dependence on the mean degree $\langle q \rangle$ in contrast to the result $J_c = 2/[\pi g(0) \langle q \rangle]$ obtained in Ref.~\cite{Lee2005} for the standard Kuramoto model on the Erd\H{o}s--R\'{e}nyi networks with a one-peaked distribution function $g(\omega)$ of natural frequencies.



\section{Kuramoto model on scale-free networks}
\label{scale-free nets}

\begin{figure}[b]
\centering
\includegraphics[width=0.48 \textwidth]{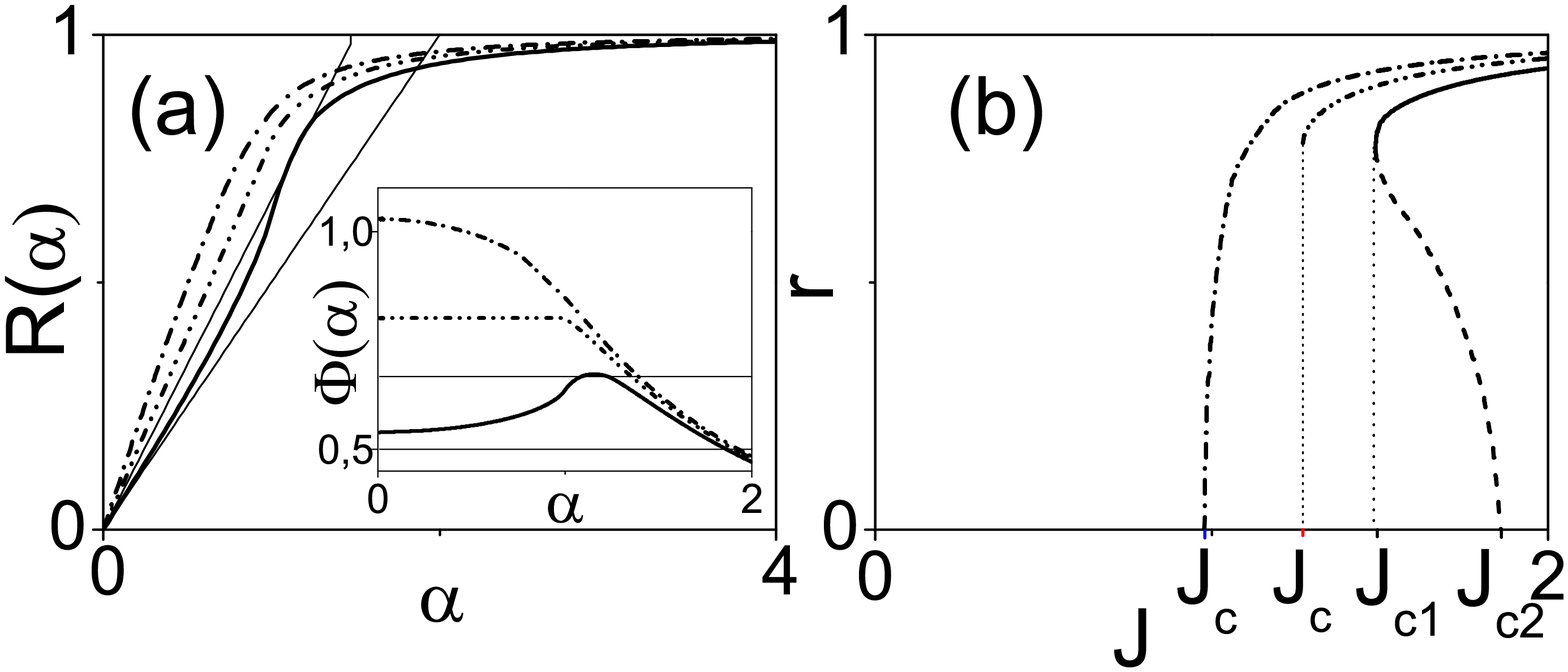}
\caption{Kuramoto model with frequency-degree correlations on scale-free networks. (a) $R(\alpha)$ versus $\alpha$ for scale-free networks with the degree distribution exponents $\gamma=3.2$, $\gamma=3$, and $\gamma=2.8$ (dash-dot, dash-dot-dot,  and solid lines from left to rigth, respectively).
The thin solid lines represent the linear function $\alpha/J$ for $J=1.5$ and $J=2$, respectively. The intersection of $R(\alpha)$ and $\alpha/J$ determines the solution of Eq.~(\ref{eq6c}) and, in turn, Eq.~(\ref{r-alpha}) gives the order parameter $r$. Inset represents the function $\Phi(\alpha)$, Eq.~(\ref{eq11a}),
for scale-free networks with the degree distribution exponents $\gamma=3.2$, $\gamma=3$, and $\gamma=2.8$ (dash-dot, dash-dot-dot, and solid lines, from up to down, respectively). The
thin solid horizontal
lines display the value of $1/J$ for $J=1.5$ and $J=2$, respectively. The intersection between the solid and dashed lines determines the order parameter $r$, Eq.~(\ref{r-alpha}). (b) The order parameter $r$ versus $\alpha$ for scale-free networks with the degree distribution exponents $\gamma=3.2$, $\gamma=3$, and $\gamma=2.8$ (dash-dot, dash-dot-dot, and solid line from left to right, respectively).
\label{graph2}}
\end{figure}
\begin{figure}[!htb]
\centering
\includegraphics[width=0.4\textwidth]{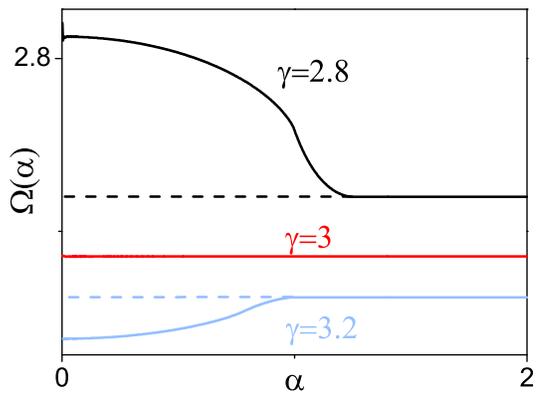}
\caption{Group angular velocity $\Omega(\alpha)$ versus $\alpha$ in the Kuramoto model with frequency-degree correlations on scale-free networks
with the degree exponent $\gamma=2.8, 3$, and $3.2$.}
\label{graph3}
\end{figure}

Let us consider the Kuramoto model with frequency-degree correlations Eq.~(\ref{linear})
on scale-free networks with a degree distribution $p(q)= A q^{-\gamma}$, where $A$ is a normalization constant and $q_0$ is the minimum degree. For this purpose we solve Eqs.~(\ref{eq6a}) and (\ref{eq6c}). The function $R(\alpha)$  given by  Eq.~(\ref{eq6b}) and our results of a numerical solution of Eqs.~(\ref{eq6a}) and (\ref{eq6c}) are represented in Fig.~\ref{graph2} at different values of the degree distribution exponent $\gamma$. Note that in this case the function $R(\alpha)$ does not depend  on the minimum degree $q_0$. Fig.~\ref{graph3} displays the function $\Omega(\alpha)$ found from a numerical solution of Eq.~(\ref{eq6c}) at different values of degree exponent $\gamma$.

Figure~\ref{graph2}(b) shows that if $\gamma > 3$ the system undergoes a second-order phase transition at $J=J_c$. At $J<J_c$, Eqs.~(\ref{eq6a}) and (\ref{eq6c}) have only a trivial solution $r=0$   that corresponds to the intersection of $R(\alpha)$ and the line $\alpha/J$ at the point $\alpha=0$.
At $J>J_c$ a non-trivial solution $r\neq 0$ emerges.
The trivial and non-trivial solutions correspond to two intersections in Fig.~\ref{graph2}(a). The solution with $r\neq 0$ is stable while the trivial one is unstable.


At $2<\gamma<3$ we find that the system undergoes a first-order transition solution at $J=J_{c1}$.
In the range $J_{c1} < J < J_{c2}$, hysteresis takes place. In this range,
there are three solutions of Eqs.~(\ref{eq6a}) and (\ref{eq6c}) [three intersections between $R(\alpha)$ and $\alpha/J$ in Fig~\ref{graph2}(a)]. The non-trivial solution with the smallest $\alpha$ is always unstable. The trivial solution $\alpha=0$ and the solution with the largest $\alpha$  correspond to stable and metastable states (this will be discussed below).

If $\gamma=3$, we find that there is a discontinuity in the order parameter $r$ at the critical coupling  but there is no hysteresis at $J> J_{c}$. It is actually a hybrid phase transition similar to the transition found for the $k$-core of random graphs \cite{dgm2006,gdm2006},
bootstrap percolation \cite{bdgm2011}, and the avalanche collapse of interdependent networks \cite{bdgm2012}. Assuming that avalanches are a
generic feature of hybrid phase transitions, we suggest that  avalanche collapse of synchronization also occurs in the Kuramoto model at the critical coupling $J_{c}$. When $J$ decreases and tends to $J_{c}$, avalanches of desynchronization of oscillators emerge
reducing the synchronization.
Namely,
when a single oscillator becomes drifting,
it triggers
an avalanche in which a large group of previously locked oscillators become drifting.
The averaged size of these avalanches approaches infinity as $J \rightarrow J_{c}$. The structure and the statistics of avalanches were studied in detail for the $k$-core problem \cite{gdm2006} and the collapse of interdependent networks \cite{bdgm2012}.

\begin{figure}[!htb]
\centering
\includegraphics[width=0.4 \textwidth]{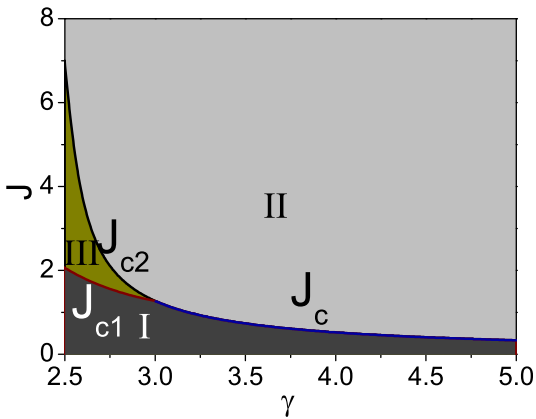}
\caption{$\gamma-J$ plane of the phase diagram of the Kuramoto model on scale-free networks with frequency-degree correlations. In region $I$ ($J < J_{c1}$) there is no spontaneous synchronization and the order parameter $r =0$. Synchronization appears in region $II$ ($J > J_{c2}$) in which the order parameter $r> 0$. Region $III$ ($J_{c1}<J<J_{c2}$) is the region of hysteresis with one metastable and one stable states.\label{graph4}}
\end{figure}

In order to study
the case of $2<\gamma\leq3$, it is convenient to rewrite Eq.~(\ref{eq6b}) in
the form
\begin{equation}
\Phi(\alpha)=1/J,\label{eq11b}
\end{equation}
where we introduced the function $\Phi(\alpha)\equiv R(\alpha)/\alpha$,
\begin{eqnarray}
&& \nonumber\Phi(\alpha)=(\gamma-2)\left(\frac{\Omega(\alpha)}{q_0}\right)^{2-\gamma} \int_{-1}^{+1} dx  {(1-\alpha x)^{\gamma-3}}\times
\\[5pt]
&&\sqrt{1-x^2}\Theta\left(1-\alpha x\right)\Theta\left(\alpha x-\frac{q_0-\Omega(\alpha)}{q_0}\right),
\label{eq11a}
\end{eqnarray}
and
the variable $x \equiv (q-\Omega(\alpha)/(\alpha q)$.
Using Eq.~(\ref{eq11b}),
we find a criterion for a first-order phase transition. Namely, a first-order phase transition  takes place if the function  $\Phi(\alpha)$ has a maximum at $\alpha \neq 0$. In order to prove this criterion,
note that $\Phi(\alpha)\rightarrow 0$
as $\alpha \rightarrow \infty$.
Therefore, for Eq.~(\ref{eq11b}) to have more than one solution, it is sufficient that $\Phi(\alpha)$ be an increasing function of $\alpha$ near $\alpha=0$.
The first derivative of $\Phi(\alpha)$ is zero at $\alpha=0$, $\left.\Phi'(\alpha=0)\equiv d \Phi(\alpha)/(d \alpha)\right|_{\alpha=0}=0$,
so
the second derivative, $\left.\Phi''(\alpha=0)\equiv\partial^2\Phi(\alpha)/(\partial^2 \alpha)\right|_{\alpha=0}$, determines the behavior of $\Phi(\alpha)$ at small $\alpha$. Then, the sufficient condition to have a first-order phase transition is $\Phi''(\alpha=0)>0$, that is
\begin{equation}
\frac{(\gamma-4)(\gamma-3)}{4(\gamma-2)}-\frac{\Omega''(0)}{\Omega(0)}>0,
\label{eq12}
\end{equation}
where $\left.\Omega''(\alpha=0)\equiv\partial^2\Omega(\alpha)/(\partial^2 \alpha)>0\right|_{\alpha=0}$. In order to find when the inequality is satisfied,
we analyze behavior of $\Omega(\alpha)$ at $\alpha=0$ for $\gamma$ close to $3$. For $|\gamma-3|\ll1$, we obtain (see appendix \ref{anexoc})
\begin{eqnarray}
&&\Omega(0)/q_0-2= \,\frac{\pi^2}{4}(\gamma-3),\label{eq15a}
\\[5pt]
&&\Omega''(0)/q_0 \simeq 1.71 \, (\gamma-3).\label{eq15b}
\end{eqnarray}
Substituting these results into Eq. (\ref{eq12}), we find that the inequality is satisfied if $\gamma<3$.
Solving numerically Eqs.~(\ref{eq6a}) and (\ref{eq6c}), we find the phase diagram shown in Fig.~\ref{graph4}. One can see that in the region $I$, there is no spontaneous synchronization. Spontaneous synchronization appears in region $II$. Region $III$ is the region with  hysteresis (there are one stable and one metastable states).


The critical behavior of the order parameter $r$ near the critical point $J_c$ can be found using the Taylor series of the function $\Phi(\alpha)$ in Eq.~(\ref{eq11b}) at the point $\alpha=0$. We find that at $\gamma>3$, the phase transition is of the second order, and the order parameter $r$ has the critical singularity (\ref{eq7}) with the critical exponent $\beta=1/2$.
At $\gamma=3$ the model undergoes a hybrid phase transition with a jump $r_c \neq 0$ of the order parameter and
demonstrates the following
critical behavior
\begin{equation}
r-r_c\propto (J-J_c)^{\beta},
\label{eq7b}
\end{equation}
where the critical exponent $\beta=2/3$ (see appendix \ref{anexoD}). The same critical exponent for the hybrid phase transition in Kuramoto model with a flat distribution of natural frequencies was found by Paz\'{o} \cite{Pazo2005}. This critical behavior is in contrast to $\beta=1/2$ found for hybrid transitions in other systems \cite{dgm2006,gdm2006,bdgm2011,bdgm2012}.
Note that in the hybrid transitions,
the distribution of avalanches over size $S$ becomes power-law at the critical point, for example, $P(S)\propto S^{-\sigma}$ with $\sigma=3/2$ for $k$-core problem. We do not know yet if the exponent  $\sigma$ takes the same value for the synchronization hybrid transition.

Thus, the analytical consideration of the Kuramoto model with frequency-degree correlations
on uncorrelated random scale-free networks shows that the type of the phase transition is changed at $\gamma=3$ from the second-order transition at $\gamma >3$ to the first-order transition at $\gamma <3$. Below we will show that simulations of the model on the static model of scale-free networks \cite{Goh1,lgkk2004,lgkk2006} confirm this analytical result. This conclusion contrasts with results of numerical simulations of G\'{o}mez-Garde\~nes et al. in Ref.~\cite{Gomez_2011} in which the
the first-order phase transition was observed even in the configuration model of scale-free networks with $\gamma \approx 3.3$.
The reason of this disagreement
may be related to the fact that G\'{o}mes et al. simulated the Kuramoto model on the top of
scale-free networks with $N=1000$ oscillators, while we simulated the Kuramoto model on networks of larger size, $N=10^4$. It is well-known that the clustering coefficient is finite in the configuration model of finite size and decreases with increasing size, approaching zero in the infinite size limit \cite{bc2003,newman2003}. Thus, networks of small size have a larger clustering coefficient in comparison with networks of larger size.
We suggest that clustering or degree-degree correlations in complex networks
may influence the synchronization phase transition
and they may be responsible for this discrepancy.
Networks generated by the static model, which we use in our simulations, are uncorrelated and have zero clustering at $\gamma >3$ in the thermodynamic limit while weak disassortative degree-degree correlations appear at $\gamma < 3$  \cite{lgkk2006}. In general, structural correlations are significant for phase transitions in complex networks. The influence of  degree-degree correlations on the percolation transition in correlated networks was demonstrated in Ref.~\cite{gdm2008}.


\section{Comparison between the annealed network approach and simulations}
\label{simulations}


In order to check the accuracy of the annealed network approach, we carried out simulations of the Kuramoto model with frequency-degree correlations Eq.~(\ref{linear}) for the static model \cite{Goh1,lgkk2004,lgkk2006} and compared the obtained results with the numerical solution of Eqs.~(\ref{eq6a}) and (\ref{eq6c}). We solved dynamical equations (\ref{eq3}) by use of the Runge-Kutta 4th order method. In our simulations, we increased and decreased the coupling constant $J$ and let the system to relax after every change of $J$. To find a correct solution of Eqs.~(\ref{eq3}), we used a time step $\Delta t=0.001$  and a coupling step $\Delta J=0.02$. The size of the network was $N=10000$.

\begin{figure}[!htb]
\centering
\includegraphics[width=0.35 \textwidth]{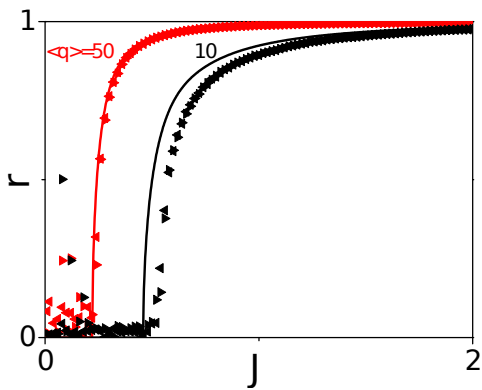}
\caption{\label{graph5} The order parameter $r$ versus the coupling $J$ for the Kuramoto model on Erd\H{o}s--R\'{e}nyi networks. Numerical simulations in the case of increasing  and decreasing $J$ are represented by the symbols ($\blacktriangleright$) and ($\blacktriangleleft$), respectively. The solid lines represent the results of the annealed network approximation [Eqs. (\ref{eq6a}) and (\ref{eq6c})]. The mean degree $\langle q \rangle=50$ and size $N=10000$.}
\end{figure}

Figure \ref{graph5} displays results of our simulations for the Erd\H{o}s--R\'{e}nyi network with the mean degree $\langle q \rangle=10$ and $\langle q \rangle=50$. One can see that the theoretical calculations and the simulations agree well if the mean degree is large enough. For $\langle q \rangle=50$ the numerical results are in good agreement with the simulation, but for $\langle q \rangle=10$ there are some differences at $J$ near $J_c$. However, even at a small mean degree  $\langle q \rangle$, $\mean{q} \lesssim 10$, the annealed network approach give us a good description of the Kuramoto model.


\begin{figure}[!htb]
\centering
\includegraphics[width=0.35 \textwidth]{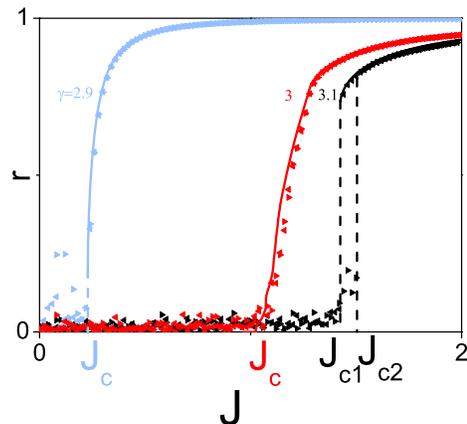}
\caption{\label{graph6} Order parameter $r$ of synchronization versus the coupling $J$ for the Kuramoto model on scale-free networks (the static model) with frequency-degree correlations. Blue, red, and black symbols  represent the results of our numerical simulations for scale-free networks with the degree distribution exponent $\gamma=2.9$, $\gamma=3$, and $\gamma=3.1$, respectively. Numerical simulations in the case of increasing  and decreasing $J$ are represented by the symbols ($\blacktriangleright$) and ($\blacktriangleleft$), respectively. The solid lines represent the results of numerical solution of Eqs.~(\ref{eq6a}) and (\ref{eq6c}). Size $N=10^4$.}
\end{figure}

Figure \ref{graph6} displays our results for a scale-free network generated by the static model \cite{Goh1} with the mean degree $\langle q \rangle=50$ and size $N=10000$. As one can see, the simulations are in a good agreement with the annealed network approach, Eqs. (\ref{eq6a}) and (\ref{eq6c}). With decreasing
mean degree, some deviations between the simulation and the annealed network approximation appear, but the type of the phase transition is the same. Note that the critical coupling $J_c$ obtained by use of the annealed network approach is
slightly smaller than $J_c$ observed in simulations.


\section{Kuramoto model on a star graph}
\label{stargraph}


In order to reveal the role of hubs in the Kuramoto model on complex networks, we study the model on star graphs. A particular case of this system has been considered in Ref.~\cite{Gomez_2011}. In this paper, the Kuramoto model was solved explicitly in the case when the central oscillator has the natural frequency equal to the number $K$ of nearest neighbors while the neighbors have the same natural frequency equal to 1. G\'{o}mez-Garde\~nes \emph{et al.} found that this model undergoes the first order phase transition at a critical coupling \cite{Gomez_2011}.
Another synchronization model, Stuart-Landau oscillators, was considered on a star graph in Ref.~\cite{Frasca:fbk12}.

Here, we obtain an exact solution of the Kuramoto model on a star graph with an arbitrary natural frequencies distribution in the limit of large number $K$ of nearest neighbors.
The dynamical equations for this model are
\begin{eqnarray}
\dot{\theta}_j=&&\omega_j+J\sum_{l=1}^{K} \sin(\theta_j-\theta_l),\label{eq16a}
\\[5pt]
\dot{\theta}_l=&&\omega_l+J\sin(\theta_j-\theta_l),\label{eq16b}
\end{eqnarray}
where Eq.~(\ref{eq16a}) is for the central node $j$, and Eq.~(\ref{eq16b}) is for its $K$ neighbors with index $l=1,2,\dots, K$. For convenience, here we define the order parameter as
\begin{equation}
re^{i\psi}\equiv \frac{1}{K}\sum_{l=1}^{K} e^{i\theta_l}.
\label{orderparameter2}
\end{equation}
Introducing this order parameter into Eqs.~(\ref{eq16a}) and (\ref{eq16b}), we obtain
\begin{eqnarray}
\dot{\theta}_j-\Omega=&&(\omega_j-\Omega)-J K r \sin(\theta_j-\psi), \label{eq17a}
\\[5pt]
\dot{\theta}_l-\dot{\theta_j}=&&(\omega_l-\dot{\theta_j})-J\sin(\theta_l-\theta_j).
\label{eq17b}
\end{eqnarray}
The central node $j$ is the leader and it must be locked in a synchronized state,
\begin{equation}
\omega_j-\Omega=J K r \sin\left(\theta_j-\psi\right).
\label{stablestate3}
\end{equation}
where $\Omega\equiv\dot{\theta}_j$ is the group angular velocity.  There are two different kinds of solutions of Eq. (\ref{eq17b}) for a steady state. If $\left.|\omega_l-\Omega|<J\right.$, the oscillator $l$ is locked. In this case, Eq.~(\ref{eq3}) has a stable solution, with $\dot{\theta_l}=\Omega$ and
\begin{equation}
\omega_l-\Omega=J \sin(\theta_j-\theta_l).
\label{stablestate2}
\end{equation}
If $|\omega_l-\Omega|>J$, the oscillator $l$ is drifting.
Contributions of locked and drifting oscillators to the order parameter $r$ can be obtained by using the method described in Appendix \ref{anexA}. Then, equation (\ref{orderparameter2})
takes a form,
\begin{eqnarray}
\nonumber
r e^{i(\psi-\theta_j)}&=&\frac{1}{K} \sum_{l=1}^{N}  e^{i(\theta_l-\theta_j)}\Theta\left(1-\left|\frac{\omega_l-\Omega}{J r}\right|\right)
\label{eq4a}
\\[5pt]
&+&\frac{1}{K} \sum_{l=1}^{N}  e^{i(\theta_l-\theta_j)}\Theta\left(\left|\frac{\omega_l-\Omega}{J r }\right|-1\right).
\label{eq4b}
\end{eqnarray}
We introduce the distribution function of the natural frequencies of the neighbors as follows
\begin{equation}
g(\omega)\equiv \sum_{l=1}^{K}\delta(\omega-\omega_l)/K
.
\label{g-omega}
\end{equation}
Separating the imaginary and real parts in Eq.~(\ref{eq4b}), in the limit $K\gg 1$, we obtain a set of two equations for $r$ and $\Omega$:
\begin{eqnarray}
&&r^2\!=\!\left(\frac{\Omega\!{-}\!\omega_j}{K J}\right)^2\!{+}\!\left[ \int_{-J}^{+J}\! d\omega \, g\left(\omega{+}\Omega\right)  \sqrt{1\!{-}\!\left(\frac{\omega}{J}\right)^2}\right]^2, \label{eq18a}
\\[5pt]
&&\nonumber \frac{\Omega-\omega_j}{K}= \int_{-\infty}^{+\infty}  d\omega \,g\left(\omega+\Omega\right)\omega \times
\\[5pt]
&& \,   \left[1-\sqrt{1-\left(\frac{J}{\omega}\right)^2}\Theta\left( |\omega|-J\right)\right].
\label{eq18b}
\end{eqnarray}
In fact, equation (\ref{eq18a}) determines $r$ as a function of the group angular velocity $\Omega$ that must be found by solving Eq.~(\ref{eq18b}).


\begin{figure}[!htb]
\centering
\includegraphics[width=0.48\textwidth]{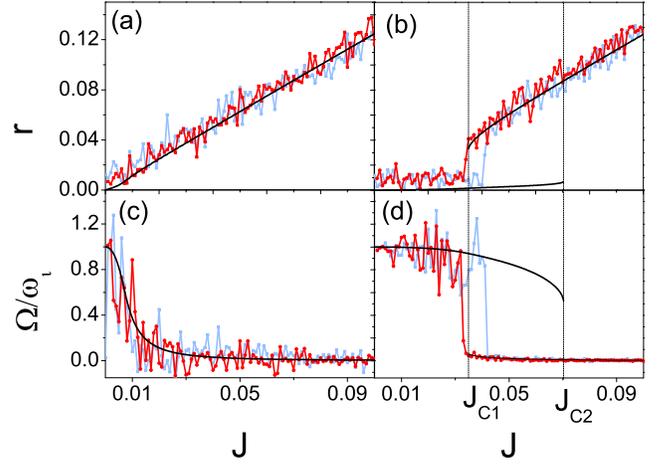}
\caption{Order parameter $r$ and the group angular velocity $\Omega$ versus the coupling $J$ in the Kuramoto model on a star graph. The results of simulations of the model for increasing  and decreasing $J$ are represented by blue and red lines with dots, respectively. We used a normal distribution of the natural frequencies  with variance ($\sigma=0.5$) and zero mean value, $\overline{\omega}=0$. The number of oscillators is $K=10000$. The central node frequency is $\omega_j=1 < \omega_c \approx 2.12$ for the panels (a) and (c), and $\omega_j=10 >\omega_c$ for the panels (b) and (d). The numerical solution of Eqs.~(\ref{eq18a}) and (\ref{eq18b}) is represented by solid lines on these panels. \label{graph7}}
\end{figure}


The analysis of Eqs. (\ref{eq18a}) and (\ref{eq18b}) shows that if the difference $\omega_j-\mean{\omega_l}$ between the natural frequency $\omega_j$ of the central oscillator and the averaged natural frequency $\mean{\omega_l}$ of its neighbors is smaller than a critical value $\omega_c$,
then synchronization between oscillators occurs at any nonzero coupling $J$. Here, $\mean{\omega_l}\equiv K^{-1}\sum_{l=1}^K \omega_l$ tends to the mean value $\overline{\omega}$ in the limit $K\rightarrow \infty$. Figure \ref{graph7}(a) shows that, in this case, the order parameter $r$ increases gradually with increasing $J$ while the group angular velocity $\Omega$ decreases.
In contrast to this case, if the difference $\omega_j-\mean{\omega_l}$ is larger than $\omega_c$, i.e.,
\begin{equation}
\omega_j-\mean{\omega_l} > \omega_c
,
\label{star1}
\end{equation}
then the Kuramoto model on the star graph undergoes a first order phase transition with hysteresis in a region $J_{c1} <J < J_{c2}$. Behavior of $r(J)$ and $\Omega(J)$ is represented in Fig. \ref{graph7}.  Near the limiting points  $J_{c1}$ and $J_{c2}$ of the metastable states, i.e., at either $0\leq J/J_{c1}-1\ll 1$ or $0\leq 1-J/J_{c2}\ll 1$,  $r$ and  $\Omega$ demonstrate a universal critical behavior:
\begin{eqnarray}
&& r - r_{c1} \propto (J/J_{c1} -1)^{1/2}, \label{lcp1-r}\\
&& r - r_{c2} \propto -(1-J/J_{c2})^{1/2}, \label{lcp2-r}\\
&& \Omega - \Omega_{c1} \propto -(J/J_{c1} -1)^{1/2}, \label{lcp1-w}\\
&& \Omega - \Omega_{c2} \propto (1-J/J_{c2})^{1/2}, \label{lcp2-w}
\end{eqnarray}
where $r_{c1}, r_{c2}, \Omega_{c1}$, and $\Omega_{c2}$ are values of $r$ and $\Omega$ in the limiting points  $J_{c1}$ and $J_{c2}$ (see Appendix~\ref{anexB}). This first-order phase transition is similar to one we found in scale-free networks in Secs.~\ref{scale-free nets} and \ref{simulations}.

At $\omega_j=\omega_c$, the order parameter $r$ and the angular group velocity $\Omega$ are  continuous functions of $J$. However, at a critical coupling $J_c$, the model has critical behavior, 
\begin{eqnarray}
&& r- r_c \propto (J/J_{c} -1)^{1/3}, \label{scp1} \\
&& \Omega- \Omega_{c} \propto -(J/J_{c} -1)^{1/3}, \label{scp2}
\end{eqnarray}
that is different from Eqs. (\ref{lcp1-r})-(\ref{lcp2-w}) (see Appendix~\ref{anexB}).

To check the analytical approach, we compare the numerical solution of Eqs. (\ref{eq18a}) and (\ref{eq18b})  with simulations of the Kuramoto model on the star graph. In simulations, we solved dynamical equations (\ref{eq16a}) and (\ref{eq16b}) by use of the Runge-Kutta 4th order method for the normal distribution function $g(\omega)$ with  the variance $\sigma=0.5$ and zero mean value $\overline{\omega}=0$.
In our simulations, we increased and decreased the coupling constant $J$ step by step and allowed the system to relax to a steady state after every step.
We used a time step $\Delta t=0.01$  and the coupling step $\Delta J=0.005$.  The number of neighbors  was $K=10000$. Figure \ref{graph7} displays our results of simulations and numerical solutions of Eqs.~(\ref{eq18a}) and (\ref{eq18b}).
One can see that despite some noise, the results of the numerical solution of Eqs. (\ref{eq18a}) and (\ref{eq18b}) and the simulations are in a good agreement.


Analyzing Eqs.~(\ref{eq18a}) and (\ref{eq18b}), we find that the critical value $\omega_c$ depends on the distribution function of natural frequencies of the oscillators around the central oscillator. In the case of the normal distribution of $\omega_l$ we obtain $\omega_c/\sigma \simeq 4.25$. The critical frequency $\omega_c$ is calculated in Appendix \ref{anexB}. Figure~\ref{graph8} displays the  $\omega_j-J$ plane of the phase diagram of the Kuramoto model on a star graph.


\begin{figure}[!htb]
\centering
\includegraphics[width=0.33 \textwidth]{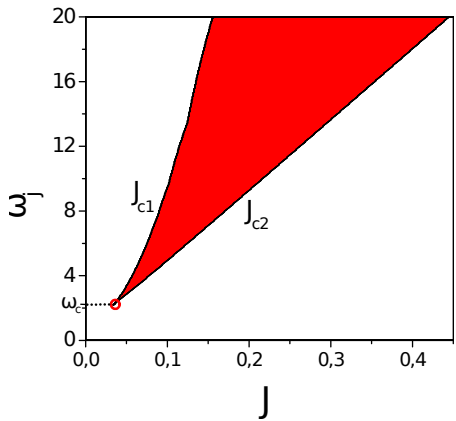}
\caption{$J-\omega_j$ plane of the phase diagram of the Kuramoto model on a star graph with $10000$ neighboring oscillators in the case of a normal distribution of the natural frequencies  with variance ($\sigma=0.5$) and zero mean value, $\overline{\omega}=0$. The gray (orange online) region represents the region with hysteresis. The 
open dot marks the critical value $\omega_c$ of the frequency $\omega_j$ of the central oscillator. The critical frequency $\omega_c \approx 2.12$
\label{graph8}}
\end{figure}


The
conclusion that the Kuramoto model on a star graph satisfying the condition Eq.~(\ref{star1}) undergoes a first-order phase transition, gives a qualitative understanding of the role of hubs in the first-order phase transition discussed in Sec.~\ref{scale-free nets}.
Indeed, in the Kuramoto model with frequency-degree correlations on a complex network, Eq.~(\ref{linear}), the natural frequency of a hub have a good chance to satisfy the condition (\ref{star1}), since a high degree of a vertex guarantees its high natural frequency.
Therefore, if the fraction of hubs is sufficiently large, then they can
induce a first-order synchronization phase transition.


\section{Conclusions}
\label{concl}


In the present paper we developed an analytical approach based on the annealed network approximation
to the Kuramoto model with
linearly coupled natural frequencies and the degrees of vertices in complex networks (frequency-degree correlations).  We demonstrated that the model undergoes a first order synchronization phase transition on uncorrelated scale-free networks with the degree distribution exponent $2<\gamma<3$, i.e., in the case of divergent second moment of degree distribution. A second-order synchronization transition occurs at $\gamma>3$, i.e., when the second moment is finite. At $\gamma=3$, the model undergoes a hybrid phase transition that combines a jump of the order parameter at the critical point as in first-order phase transitions and critical phenomena near the critical point as in second-order phase transitions. In the case of hybrid transition, avalanche collapse of synchronization  occurs at the critical coupling $J_{c}$.
We compared our analytical calculations with numerical simulations for Erd\H{o}s--R\'{e}nyi and scale-free networks of size $N=1000-10000$. Our results demonstrated that the annealed network approach is accurate if the size of the network and the mean degree are sufficiently large.
In order to understand a mechanism of the first-order synchronization phase transition, we also analyzed analytically and numerically the Kuramoto model on star graphs and showed that the central oscillator plays the role of the leader in synchronization. If the difference between a natural frequency $\omega_j$ of the central oscillator and the averaged natural frequency $\overline{\omega}$ of its neighbors is smaller than a certain critical value $\omega_c$, i.e., $\omega_j-\overline{\omega} <\omega_c$, then synchronization occurs at any nonzero coupling $J$ and it is gradually enhanced with increasing $J$. In contrast to this case, if $\omega_j-\overline{\omega} <\omega_c$,
then the system undergoes a first-order transition into a synchronized state. In this case, hysteresis takes place in a certain range of the coupling $J$. This result evidences that hubs in a complex network of phase oscillators may
play a role of driving force for
a first-order phase transition.



\begin{acknowledgments}
This work was partially supported by the FCT
projects PTDC:
FIS/71551/2006, FIS/108476/2008, SAU-NEU/103904/2008, MAT/114515/2009,  PEst-C/CTM/LA0025/2011, and FET IP Project MULTIPLEX 317532.
\end{acknowledgments}


\appendix


\section{Contribution of drifting oscillators to the order parameter}
\label{anexA}


Let us calculate the contribution of drifting oscillators to the order parameter $r$ in Eq.~(\ref{eq4}). Despite a random
movement of these oscillators, the total contribution of drifting oscillators to the order parameter becomes time independent in the limit $t\rightarrow \infty$ and $N \rightarrow \infty$. This contribution can be calculated by use of the density function $\rho(\dot{\theta},qJr)$ that measures the density of oscillators with the angular velocity $\dot{\theta}$ and the mean degree $q$ as it was shown in Ref.~\cite{bu2008}. In this Appendix we present a new method that can also be useful for studying dynamics and relaxation in the Kuramoto model. For simplicity, we calculate this contribution for star graphs (see Sec. \ref{stargraph}), but the method can be generalized
to other graphs.
We begin with an analytic solution of Eq.~(\ref{eq16b}) for drifting oscillators with natural frequencies satisfying the inequality $|\omega_j-\Omega |> J$. This solution is
\begin{widetext}
\begin{equation}
\theta_l-\theta_j=2 \arctan\left\{
\frac{J+\tan\left[
(k_l+t)
\sqrt{(\omega_l-\Omega)^2-J^2}/2 \right]\sqrt{(\omega_l-\dot{\Psi})^2-J^2}}{\omega_l-\Omega} \right\},
\label{eqanex1}
\end{equation}
\end{widetext}
where $k_j$ are parameters determined by initial conditions at given $\omega_l$ and $J$ defined in the interval $0<k_j<2\pi$. Substituting Eq.~(\ref{eqanex1}) into the last term in Eq. (\ref{eq4}), we obtain that in the thermodynamic limit the contribution of all drifting oscillators is given by the equation
\begin{widetext}
\begin{equation}
\int_{-\infty}^{+\infty}\int_{-\infty}^{+\infty} dk d\omega \, G(k, \omega+\Omega)  \exp\!\left\{2i\, \arctan\!\left[\left(J + \tan\left( \frac{k+t}{2}\sqrt{w^2 - J^2} \right) \sqrt{w^2 - J^2}\right) w^{-1} \right]\right\}\Theta(|\omega|-J),
\label{eqanex2}
\end{equation}
\end{widetext}
where we introduced the function
\begin{equation}
G(k,\omega+\Omega)\equiv \frac{1}{K} \sum^{K}_{j=1} \delta(\omega+\Omega-\omega_j)\delta(k-k_j).
\end{equation}
In order to simplify the calculations, we introduce a variable $a\equiv \omega \sqrt{1-\left(J/\omega\right)^2}$. Then Eq.~(\ref{eqanex2}) takes the form
\begin{widetext}
\begin{equation}
\int_{-\infty}^{+\infty}\int_{-\infty}^{+\infty} dk da \,  \frac{G\left(k, \omega(a)+\Omega\right)}{\sqrt{1+\left(J/a \right)^2}}  \exp\!\left\{2i\, \arctan\!\left[\frac{J + \tan\left( \frac{1}{2}(k+t)\, |a|\right) |a|}{\sqrt{1+\left(J/a \right)^2}}\right]\right\}.
\label{eqanex3}
\end{equation}
\end{widetext}
In order to find this integral we replace the variable of integration $a$ to $a=a_n/t+2y/t$, where $a_n$ is a discrete variable given by $a_n=2 n\pi$ with $\, n\in \mathbb{Z}$, and $y$ is a continuous variable defined in the interval $[0,\pi]$. Then Eq.~(\ref{eqanex3}) takes the form
 \begin{widetext}
 \begin{equation}
\frac{2}{t}\int_{-\infty}^{+\infty} \sum_n \int_{0}^{ \pi} dk dy \, \frac{G\left(k, \omega(a_n/t+2y/t)+\Omega\right)}{\sqrt{1+\left(J/a \right)^2}}   \exp\!\left\{2i\, \arctan\!\left[\frac{J + \tan\left(|a_n|\frac{k+t}{2t} +y\frac{k+t}{t}\right)\left( |a_n|/t+2y/t\right)}{\sqrt{1+\left(\frac{J}{a_n/t+2y/t}\right)^2}}\right]\right\}.
\label{eqanex4}
\end{equation}
\end{widetext}
Since $0<y<\pi$ and $0<k<2\pi$, in the limit $t\rightarrow +\infty$, we have  $k/t\rightarrow 0$ and $y/t\rightarrow 0$.  	Furthermore, the summation over $a_n$ can be represented as integration over $a$. The resulting function is a periodic function of $y$. The integration over $y$ remove  the dependence on the initial conditions and we obtain a triple integral that does not depend on time and the initial conditions. Introducing a function $\left. g(\omega+\Omega)\equiv \int_{-\infty}^{\infty} G(k,\omega+\Omega)dk\right.$ we obtain a well-known result for the contribution of drifting oscillators to the order parameter $r$ in Eq.~(\ref{eq4}),
\begin{equation}
i\int_{-\infty}^{\infty} d\omega\, g(\omega+\Omega)\frac{\omega}{J}[1-\sqrt{1-(J/\omega)^2}] \, \Theta(|\omega|-J).
\label{eqanex6}
\end{equation}


\section{Analysis of the group velocity function
}
\label{anexoc}


In the case of the Kuramoto model with frequency-degree correlations on scale-free networks with degree distribution $p(q) \propto q^{-\gamma}$ at $q \geq q_0$, the group velocity function $\Omega(\alpha)$ in Eq.~(\ref{eq6a}) has no explicit expression. Nevertheless, it is possible to find its asymptotic behavior if the degree distribution exponent $\gamma$ is close to $3$. In this case, Eq.~(\ref{eq6a}) takes the form
\begin{eqnarray}
\nonumber &&\mean{q}-\Omega(\alpha)=\left(\gamma-1\right)q_0^{\gamma-1}\int_{q_0}^{+\infty} dq \, q^{-\gamma} (q-\Omega(\alpha)) \times \\&& \sqrt{1-\left(\frac{\alpha q }{q-\Omega(\alpha)}\right)^2}\Theta \left(1-\left|\frac{q \alpha}{q-\Omega(\alpha)}\right|\right).
\label{eqanexc1}
\end{eqnarray}
Using a new variable of integration, $x\equiv q/(q-\Omega)$, we can rewrite this equation as
\begin{eqnarray}
\nonumber
&&\frac{\Omega(\alpha)^{\gamma-2}}{q_0^{\gamma\!{-}\!1}}\!\left(\frac{q_0}{\gamma\!{-}\!2}{-}\frac{\Omega(\alpha)}{\gamma\!{-}\!1}\right) =\!{-}\!\left(\int_{-B}^{-\infty} dx\!+\!\int_{+\infty}^{1}dx\right)\, \times \\
&&x^{-3} \left(\frac{x}{x-1}\right)^{\gamma-3} \sqrt{1-(\alpha x)^2}\Theta \left(1-|\alpha x|\right)
,
\label{eqanexc2}
\end{eqnarray}
where $B \equiv 1/(\Omega(\alpha)/q_0{-}1)$. At $\gamma=3$, this equation takes the simple form:
\begin{eqnarray}
\nonumber
&&\frac{2{q_0}{-}\Omega(\alpha)}{2{q_0}}=-\left(\int_{B}^{+\infty}\!dx\!-\int_{1}^{+\infty}\!dx\right)\times
\\[5pt]
&&x^{-3}\sqrt{1{-}(\alpha x)^2}\Theta \left(\!1-|\alpha x|\right).
\label{eqanexc3}
\end{eqnarray}
This equation has a solution $\Omega=2 q_0$ for any $\alpha$.

\begin{figure}[!htb]
\centering
\includegraphics[width=0.35 \textwidth]{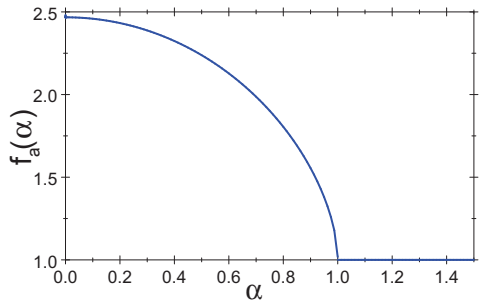}
\caption{Function $f_a(\alpha)$
from Eq.~(\ref{eqanexc6}).
\label{graph9}}
\end{figure}

If $\gamma$ is close to 3, i.e., $\gamma=3+\delta$ where $|\delta| \ll 1$, then we look for a solution of Eq.~(\ref{eqanexc2}) in the form
\begin{equation}
\Omega(\alpha)=q_0 [2 + \Delta]
\label{eqanexc4}
\end{equation}
with $|\Delta|<<1 $. We find
\begin{equation}
\Delta=-\delta f_a(\alpha),
\label{eqanexc5}
\end{equation}
where the function $f_a(\alpha)$ is defined as follows,
\begin{equation}
f_{a}(\alpha){\equiv}\frac{1\!{-}\!\int_1^{+\infty}\!
\displaystyle{\frac{dx}{x^3}\ln\left(\!\frac{\!x{+}1}{\!x{-}1\!}\!\right)}
\sqrt{\!1{-}\alpha^2 x^2} \,\Theta(\!1{-}|\alpha x|\!)}{1{-}\sqrt{1{-}\alpha^2}\,\Theta(1{-}|\alpha|)}.
\label{eqanexc6}
\end{equation}
Figure~\ref{graph9} displays the function $f_a(\alpha)$.
This function determines the behavior of the group velocity function $\Omega(\alpha)$.


\section{Critical exponent for the hybrid phase transition}
\label{anexoD}


Let us study the critical behavior of the order parameter $r$ of the Kuramoto model with frequency-degree correlations on scale-free networks with the degree distribution exponent $\gamma=3$. Using the solution $\Omega=2 q_0$ obtained in Appendix \ref{anexoc}, we find that the function $\Phi(\alpha)$ in Eq.~(\ref{eq11a}) takes the form
\begin{equation}
\Phi(\alpha)=\frac{1}{2} \int_{-1}^{+1} dx \sqrt{1-x^2}\,\Theta \left(1-|\alpha x|\right).
\label{eqanexd1}
\end{equation}
Solving Eq.~(\ref{eq11a}), we find that $\alpha=1$ at the critical point $J=J_c$. The region $\alpha>1$ corresponds to the synchronized state, and
\begin{equation}
\Phi(\alpha)= \int_{0}^{+1/\alpha} dx \sqrt{1-x^2}.
\label{eqanexd2}
\end{equation}
Near the critical point when $\delta\alpha\equiv\alpha-1 \ll 1$,
we obtain $\Phi(\alpha)$ in the leading order in $\delta\alpha$,
\begin{equation}
\Phi(1+\delta\alpha)=\frac{\pi}{4} - \frac{2\sqrt{2}}{3}(\delta\alpha)^{3/2}.
\label{eqanexd4}
\end{equation}
Expanding Eq.~(\ref{eq11b}) in a Taylor series in $J-J_c$, and using $\Phi(\alpha)$ from Eq.~(\ref{eqanexd2}), we find the critical behavior of the order parameter $r$ above the critical coupling $J_{c}$ of the hybrid phase transition, i.e., at $J > J_{c}$,
\begin{equation}
r-r_c\propto\left(J-J_{c}\right)^{\beta}.
\label{eqanexd3}
\end{equation}
Here the critical coupling is $J_c = 4/\pi$, the jump of the order parameter is $r_c=1/J_c=\pi/4$, and the critical exponent is $\beta=2/3$.



\section{Analytical analysis of the Kuramoto model on a star graph}
\label{anexB}


In the case of the Kuramoto model on a star graph, we analyze Eq.~(\ref{eq18b}) to determine the group angular velocity $\Omega$ as a function of the coupling $J$. Note that the right-hand side of Eq.~(\ref{eq18b}) tends to zero with increasing $K$. Therefore, at $K \gg 1$, a solution  $\Omega(J)$ is small at small $J$ and can be written as a Taylor series
in $J$.
In the leading order in $J$, equation (\ref{eq18b}) takes a form
\begin{equation}
\frac{\Omega-\omega_j}{K}= A\left(\Omega\right)\, J^2+O(J^4)+\cdots, \label{eqanexb1a}
\end{equation}
where we introduced
\begin{equation}
A\left(\Omega\right)\equiv(1/2)\int_{-\infty}^{+\infty} \frac{d\omega}{\omega} \, g\left(\omega+\Omega\right).\label{eqanexb1b}
\end{equation}
One notes, that at the limiting points $J_{c1}$ and $J_{c2}$ of the metastable states, the first derivatives of the right- and left-hand sides of Eq.~(\ref{eqanexb1a})  with respect to $\Omega$ becomes equal. This leads to an equation
\begin{equation}
1= KJ_{c1(2)}^2 A'\left(\Omega_{c1(2)}\right),
\label{eqanexb1c}
\end{equation}
where $A'(\Omega)\equiv dA(\Omega)/d\Omega$. Thus, the group angular velocities $\Omega_{c1}$ and $\Omega_{c2}$ and the critical couplings $J_{c1}$ and $J_{c2}$ are determined by Eqs.~(\ref{eqanexb1a}) and (\ref{eqanexb1c}). Substituting Eq.~(\ref{eqanexb1c}) into Eq.~(\ref{eqanexb1a}), we obtain an equation for $\Omega_{c1(2)}$,
\begin{equation}
A'\left(\Omega_{c1(2)}\right)\left(\Omega_{c1(2)}-\omega_j \right)= A\left(\Omega_{c1(2)}\right), \label{eqanexb2a}
\end{equation}
From Eq.~(\ref{eqanexb1c}), we find
\begin{equation}
J_{c1(2)}\propto 1/\sqrt{K}.\label{couplingwithK}
\end{equation}
Solving Eq.~(\ref{eqanexb1a}) near the limiting points  $J_{c1}$ and $J_{c2}$ of the metastable states, i.e., at either $0\leq J/J_{c1}-1\ll 1$ or $0\leq 1-J/J_{c2}\ll 1$, we find
\begin{eqnarray}
&&\Omega(J)=\Omega_{c1}-B_1 (J/J_{c1} -1)^{1/2},
\label{crit-1}\\
&&\Omega(J)=\Omega_{c2}+B_2 (1-J/J_{c2})^{1/2},
\label{crit-2}
\end{eqnarray}
where $B_{1(2)}=2|A(\Omega_{c1(2)})/A''\left(\Omega_{c1(2)}\right)|^{1/2}$. According to Eq.~(\ref{eq18a}), the order parameter $r$ also has this kind of singular behavior near $J_{c1}$ and $J_{c2}$.

Now let us find the critical frequency $\omega_c$ of the central oscillator at which hysteresis disappears, i.e., $J_{c1}=J_{c2}=J_c$.
Analyzing Eq.~(\ref{eqanexb1a}), we find that $J_c$, $\Omega_c$, and $\omega_c$ can be found from a set of equations
\begin{eqnarray}
&& \Omega_c-\omega_c= KJ_{c}^2 A\left(\Omega_c\right), \nonumber \\
&& 1=K J_{c}^2 A'\left(\Omega_{c}\right), \label{crit-1}\\
&& A''\left(\Omega_{c}\right)=0. \nonumber
\end{eqnarray}
For the gaussian distribution with variance $\sigma$ and zero mean value we find $\omega_c/\sigma \simeq 4.25$.  In order to find critical behavior near this special point, $0\leq |1-J/J_{c}|\ll 1$, we solve Eq.~(\ref{eqanexb1a}) and find
\begin{equation}
\Omega(J)=\Omega_{c}-B_3 (J/J_{c1} -1)^{1/3}, \label{crit-3}
\end{equation}
where $B_{3}=|12A(\Omega_{c})/A'''\left(\Omega_{c}\right)|^{1/3}$. Thus, at $\omega_j=\omega_c$, the order parameter $r$ and the angular group velocity are a continuous function of $J$ but, at $J=J_c$, they have a singular behavior Eq.~(\ref{crit-3}).


\begin{thebibliography}{99}

\bibitem{Pikovsky}
A.~Pikovsky, M.~Rosenblum, and J.~Kurths, \textit{Synchronization: A Universal Concept in Nonlinear Sciences} (Cambridge University Press, Cambridge, 2001).

\bibitem{Kuramoto75} Y. Kuramoto, in \emph{International Symposium on Mathematical Problems in Theoretical Physics}, H. Araki, ed. Lecture Notes in Physics 39 (Springer, New York, 1975).

\bibitem{Kuramoto84} Y. Kuramoto, \emph{Chemical Oscillations, Waves, and Turbulence} (Springer, Berlin, 1984), pp. 68-77.

\bibitem{arenas}
A.~Arenas, A.~D\'{\i}az-Guilera, J.~Kurths, Y.~Moreno, and C.~Zhou, Phys. Rep. \textbf{469}, 93 (2008).

\bibitem{Acebron2005}
J.~A.~Acebr\'{o}n, L~.L.~Bonilla, C~.J.~P\'{e}rez Vicente, F.~Ritort, and R.~Spigler, Rev. Mod. Phys. \textbf{77}, 137 (2005).

\bibitem{Dorogovtsev:dgm08}
S.~N.~Dorogovtsev, A.~V.~Goltsev, and J.~F.~F.~Mendes, Rev. Mod. Phys. \textbf{80}, 1275 (2008).


\bibitem{Strogatz1991}
S.~H.~Strogatz and R.~E.~Mirollo, J. Stat. Phys. \textbf{63}(3-4), 613 (1991).




\bibitem{Pazo2005}
D.~Paz\'{o}, Phys. Rev. E \textbf{72}, 046211 (2005).


\bibitem{bu2007}
L.~Basnarkov and V.~Urumov, Phys. Rev. E \textbf{76}, 057201 (2007).


\bibitem{bu2008}
L.~Basnarkov and V.~Urumov, Phys. Rev. E \textbf{78}, 011113 (2008).

\bibitem{ab2001}  R.~Albert and A.{-}L.~Barab\'{a}si, Rev.~Mod.~Phys. {\bf 74},
47 (2002).

\bibitem{dg2002} S.~N.~ Dorogovtsev and J.~F.~F.~ Mendes, Adv. Phys. \textbf{51}, 1079 (2002).

\bibitem{newman2003}M.~E.~J.~Newman, SIAM Review \textbf{45}, 167 (2003).


\bibitem{Ichinomiya2004}
T.~Ichinomiya, Phys. Rev. E \textbf{70}, 026116 (2004).

\bibitem{Ichinomiya2005} T.~Ichinomiya, Phys. Rev. E \textbf{72}, 016109 (2005).

\bibitem{Lee2005} D.-S.~Lee, Phys. Rev. E \textbf{72}, 026208 (2005).

\bibitem{dgm2002}
S.~N.~Dorogovtsev, A.~V.~Goltsev, and J.~F.~F. Mendes, Phys. Rev. E \textbf{66}, 016104  (2002)


\bibitem{dgm2004}
S.~N.~Dorogovtsev, A.~V.~Goltsev, and J.~F.~F. Mendes, Eur. Phys. J. B \textbf{38}, 177 (2004)



\bibitem{Gomez_2011} J. G\'{o}mez-Garde\~nes, S.~G\'{o}mez, A.~Arenas, and Y.~Moreno, Phys. Rev. Lett. \textbf{106}, 128701 (2011).

\bibitem{gm2006} J. G\'{o}mez-Garde\~nes and Y. Moreno, Phys. Rev. E \textbf{73}, 056124
(2006).

\bibitem{Leyva2012}
I.~Leyva, R.~Sevilla-Escoboza, J.~M.~Buld\'u, I.~Sendi\~na-Nadal, J.~G\'{o}mez-Garde\~nes, A.~Arenas, Y.~Moreno, S.~G\'{o}mez, R. Jaimes-Re\'{a}tegui, and S.~Boccaletti, Phys. Rev. Lett. \textbf{108}, 168702 (2012)

\bibitem{Bianconi}
G.~Bianconi, Phys. Lett. A \textbf{303}, 166 (2002).






\bibitem{Smet_2007}
F.~Smet and D.~Aeyels, Physica D \textbf{234}, 81 (2007).



\bibitem{Goh1}
K.~-I. Goh, B.~Kahng, and D.~Kim, Phys. Rev. Lett. \textbf{87}, 278701 (2001).

\bibitem{lgkk2004} D.-S.~Lee, K.-I.~Goh, B.~Kahng, and D.~Kim, Nucl. Phys. B
696, 351 (2004).

\bibitem{lgkk2006}
J. S. Lee, K. I. Goh, B. Kahng, and D. Kim, Eur. Phys. J. B \textbf{49}, 231 (2006).

\bibitem{dgm2006} S. N. Dorogovtsev, A. V. Goltsev, and J. F. F. Mendes,
Phys. Rev. Lett. \textbf{96}, 040601 (2006).


\bibitem{gdm2006}
A.~V.~Goltsev, S.~N.~Dorogovtsev,  and J.~F.~F. Mendes, Phys. Rev. E \textbf{73}, 056101  (2006).

\bibitem{bdgm2011} G. J. Baxter, S. N. Dorogovtsev, A. V. Goltsev, and
J. F. F. Mendes, Phys. Rev. E \textbf{83}, 051134 (2011).

\bibitem{bdgm2012} G. J. Baxter, S. N. Dorogovtsev, A. V. Goltsev, and J. F. F. Mendes, e-print
arXiv:1207.0448; Phys. Rev. Lett. 109, ... (2012), in press.

\bibitem{bc2003} G. Bianconi and A. Capocci,  Phys. Rev. Lett. \textbf{90},
078701 (2003).

\bibitem{newman2003} M. E. J. Newman, Phys. Rev. E \textbf{68}, 026121 (2003).


\bibitem{gdm2008} A. V. Goltsev, S. N. Dorogovtsev, and J.~F.~F. Mendes,
Phys. Rev. E \textbf{78},  051105 (2008).

\bibitem{Frasca:fbk12}
M.~Frasca, A.~Bergner, J.~Kurths, and L.~Fortuna,
Int. J. Bifurcation and Chaos {\bf 22}, 1250173 (2012).


%

\end{thebibliography}
\end{document}